\documentclass[10pt,conference]{IEEEtran}
\usepackage{cite}
\usepackage{amsmath,amssymb,amsfonts}
\usepackage{algorithmic}
\usepackage{graphicx}
\usepackage{textcomp}
\usepackage{xcolor}
\usepackage[normalem]{ulem}

\usepackage{paralist}
\usepackage{todonotes}
\usepackage{listings}
\usepackage{subfigure}
\usepackage[hidelinks]{hyperref}

\usepackage[final]{changes}
\definechangesauthor[name="Thomas"]{TN}  
\definechangesauthor[name="Amir"]{AS}    
\definechangesauthor[name="In-Saeng"]{IS}
\setcommentmarkup{~\textsubscript{~~ #1 ~~}}
\lstdefinestyle{bourne-style}{
     captionpos=b,                                 
     basicstyle=\footnotesize\ttfamily\scriptsize, 
     keywords={},                                  
}

\IEEEoverridecommandlockouts

\begin{document}

\title{A Framework for Integrating Quantum Simulation and High Performance Computing\thanks{\footnotesize Notice: This manuscript has been
    authored in part by UT-Battelle, LLC under Contract No.
    DE-AC05-00OR22725 with the U.S.  Department of Energy. The United States
    Government retains and the publisher, by accepting the article for
    publication, acknowledges that the United States Government retains a
    non-exclusive, paid-up, irrevocable, world-wide license to publish or
    reproduce the published form of this manuscript, or allow others to do
    so, for United States Government purposes. The Department of Energy will
    provide public access to these results of federally sponsored research
    in accordance with the DOE Public Access Plan
    (http://energy.gov/downloads/doe-public-access-plan)}
}

\author{
\IEEEauthorblockN{Amir Shehata}
\IEEEauthorblockA{\textit{Oak Ridge National Laboratory}\\
                  Oak Ridge, TN, USA\\
                  shehataa@ornl.gov}
\and
\IEEEauthorblockN{Thomas Naughton}
\IEEEauthorblockA{\textit{Oak Ridge National Laboratory}\\
                  Oak Ridge, TN, USA\\
                  naughtont@ornl.gov}
\and
\IEEEauthorblockN{In-Saeng Suh}
\IEEEauthorblockA{\textit{Oak Ridge National Laboratory}\\
                  Oak Ridge, TN, USA\\
                  suhi@ornl.gov}
}


\date{}
\maketitle

\begin{abstract}

Scientific applications are starting to explore the viability of quantum computing.  
This exploration typically begins with quantum simulations that can run on existing 
classical platforms, albeit without the performance advantages of real quantum resources. 
In the context of high-performance computing (HPC), the incorporation of simulation software
can often take advantage of the powerful resources to help scale-up the simulation size.   
The configuration, installation and operation of these quantum simulation packages on
HPC resources can often be rather daunting and increases friction for experimentation by
scientific application developers. 
We describe a framework
to help streamline access to quantum simulation software running on HPC resources.
This includes an interface for circuit-based quantum computing tasks, as well as the
necessary resource management infrastructure to make effective use of the underlying
HPC resources. The primary contributions of this work include a classification of
different usage models for quantum simulation in an HPC context, a review of the software architecture
for our approach and a detailed description of the prototype implementation to experiment
with these ideas \added{using two different simulators (TNQVM \& NWQ-Sim)}.  We include initial experimental results running on the
\emph{Frontier} supercomputer at
the Oak Ridge Leadership Computing Facility~(OLCF) using a synthetic workload generated
via the SupermarQ quantum benchmarking framework.
\end{abstract}


\section{Introduction}
\label{sec:intro}

In recent years, the field of quantum computing has shown remarkable advancements, demonstrating its potential to revolutionize certain types of algorithms and applications. While quantum computing holds great promise for solving specific problems exponentially faster than classical computers, its widespread adoption for general computing remains a future prospect. In the foreseeable future, quantum computing is anticipated to coexist and collaborate with classical High-Performance Computing~(HPC) environments to harness its unique advantages \cite{8123664,10.1117/12.2303824,9537178}.

In the landscape of quantum computing, Noisy Intermediate-Scale Quantum (NISQ) \cite{Preskill2018} devices have emerged as powerful yet have limited qubit coherence times and are error-prone computational platforms. These devices, featuring a limited number of qubits, qubit coherency issues and susceptibility to environmental noise, present challenges in ensuring the correctness of quantum computations. As quantum algorithms become more intricate, the need for error correction becomes pronounced, introducing complexity and demanding sophisticated algorithms to maintain accuracy. 
Quantum simulators \cite{ALEXEEV2024666} play a pivotal role in providing researchers the ability to develop, test and debug quantum algorithms in a controlled and error-free environment.  Additionally, simulation improves availability while physical resources are in short supply.

Given the benefits of quantum simulator, this paper
provides an overview of work that seeks to streamline their use in
high-performance computing environments.  A critical component of this
process is the development of a resource management framework \cite{ALEXEEV2024666,Cicconetti_2023} that can
reduce deployment complexity and increase runtime performance/efficiency.
This framework is an important building block in the integration of HPC and
quantum computing software.

\section{HPC Integration}
\label{sec:hpcmodes}

The integration of quantum computing into HPC environments represents a
strategic approach for unlocking the potential of quantum algorithms while
maintaining the reliability and versatility of classical computing \cite{qchpc,elsharkawy2023integration}.
Integrating quantum computing into HPC ecosystems creates a symbiotic
relationship, where classical systems handle traditional tasks, and quantum
processors address specific problems for which they are uniquely suited.

\begin{figure}
    \centering
    \includegraphics[width=0.75\columnwidth]{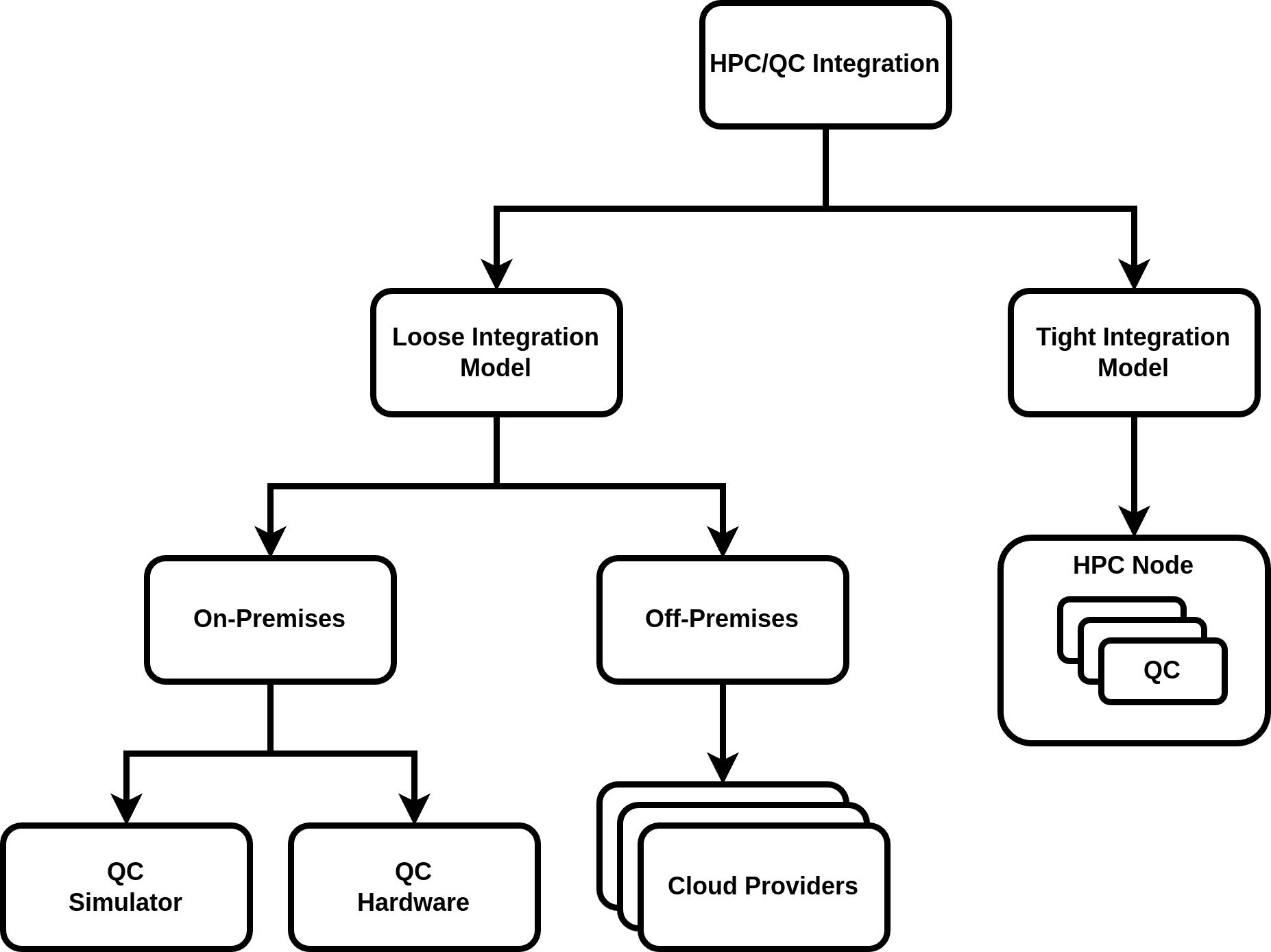}
    \caption{Integration Space}
    \label{fig:fig_integ_space}
\end{figure}
The HPC/QC integration (Figure~\ref{fig:fig_integ_space}) can be broken into the \emph{loose} and the \emph{tight integration spaces} \cite{qchpc}. The latter involves the direct integration of quantum processing units~(QPUs) into HPC nodes, similar to graphics processing units~(GPUs) in HPC compute nodes.

Our focus lies within the loose integration paradigm, where quantum computers operate as distinct entities integrated into the broader HPC environment. This loose integration can be further decomposed into  \emph{off-premises}, where Quantum resources reside in remote cloud environments, and \emph{on-premises}, which constitutes the core of our work. In the on-premises scenario, a quantum machine is integrated into the HPC center and connected to classical HPC systems via high-bandwidth interconnects and a distributed file system \cite{ALEXEEV2024666}. This allows hybrid applications to accelerate workloads which require quantum simulation.

The focus of our work revolves around the development of a framework designed for the specific needs of on-premises setting. It offers a cohesive solution for
hybrid applications to maximize the benefit HPC provides for quantum simulation. At the same time the APIs and architecture is specifically meant to allow hybrid applications to transition to real quantum hardware with
\replaced{low}{no} effort.

\subsection{Integration As It Relates to Simulations}

In the current phase of quantum hardware development, classical quantum
simulators play a crucial role in enabling researchers to develop algorithms
without being hindered by current hardware limitations.
This emphasizes the
necessity for a framework that offers a flexible infrastructure on the
\emph{Frontier} supercomputer~\cite{frontier:sc23}. This framework aims to:
\begin{compactitem}
    \item Enable researchers to utilize any quantum circuit building tool
        relevant to their problem domain.
    \item Facilitate experimentation with various simulator types.
    \item Minimize the deployment overhead for hybrid applications on the
        \emph{Frontier} supercomputer.
\end{compactitem}

\subsection{Hybrid Application Usage Patterns}
\label{sec:HybridAppUsagePatterns}

In order to develop a flexible framework which supports users requirements,
we need to examine hybrid applications usage patterns.  Hybrid HPC/QC
applications typically involve classical logic intertwined with quantum
logic. Quantum Tasks (QT) will need to be built, compiled, optimized and run
on either real quantum hardware or a quantum simulator; henceforth referred
to as quantum platform. In some cases one or multiple quantum tasks may be run
iteratively to collect a probabilistic result. These usage patterns are referred to as
\emph{single circuit} and \emph{ensemble circuit processing}, respectively. 
Subsequent quantum tasks can incorporate results from previous processing steps
to fine-tune later circuit formulation.
An additional pattern, referred to as \emph{in-sequence processing}, involves applications that require low-latency communication and control over the quantum platform for the successful execution of the task, e.g., error correction algorithms. 

\subsection{HPC/QC Integration Models}
\label{sec:HpcQcIntModels}

The current state-of-practice for integrating quantum simulators in HPC
environments is focused on pairing the simulator with processes on compute
nodes.  This \emph{per-process quantum simulator} model
has each HPC node run one
\replaced{or}{ore} more instances of
a quantum simulator, which emulates dedicated QPU resources on the local node.
This parallels the existing approach where multiple processes can run on a
single node and have access to the integrated GPUs for acceleration.

The \emph{per-process quantum simulator} model benefits small circuit
execution, as each node (or MPI process) has a dedicated simulator.
However, there can be several disadvantages to this approach, including:
 compute node over-subscription, compute node under utilization,  and
 lack of large circuit parallelization.

\begin{figure}
  \centering
    \subfigure[Single]{\label{fig:qc_res:single}\includegraphics[width=0.45\columnwidth]{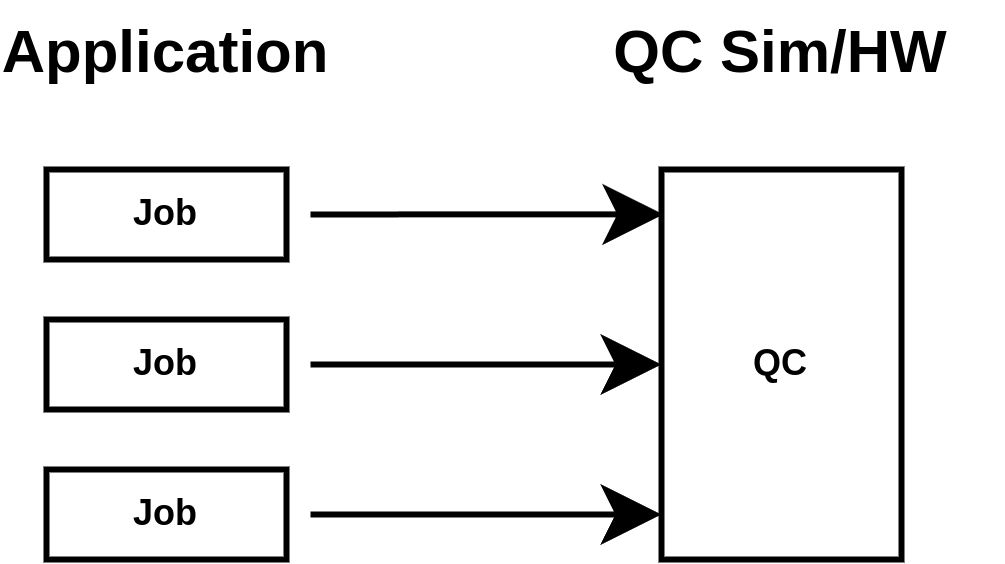}}
    \subfigure[Per-job]{\label{fig:qc_res:per_job}\includegraphics[width=0.45\columnwidth]{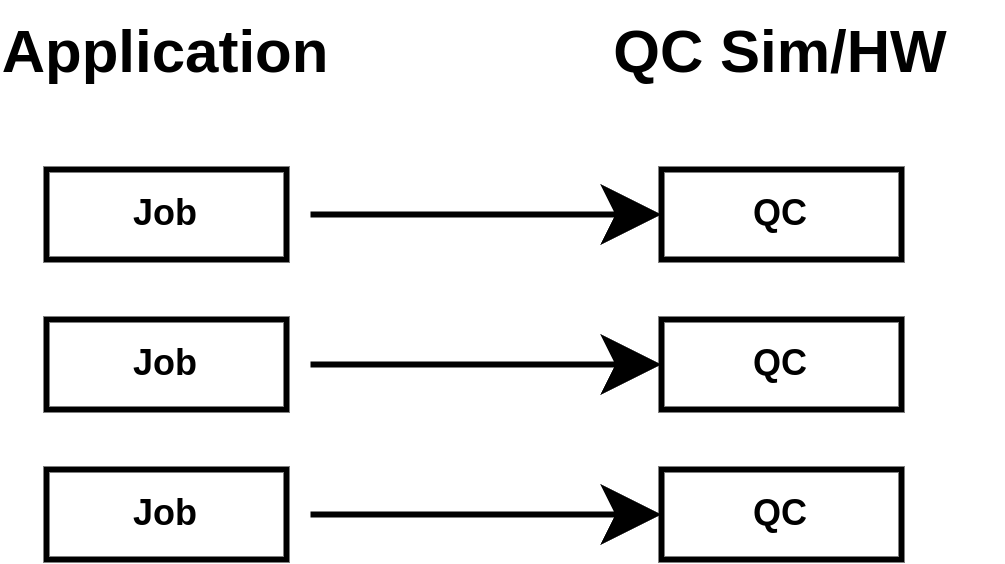}}
  \caption{Quantum Computing Resource Modes}
  \label{fig:qc_res}
\end{figure}

Given the state-of-practice integration model and its disadvantages, along
with the usage patterns outlined in
Section~\ref{sec:HybridAppUsagePatterns}, we envision two new models of
HPC/QC integration:
\emph{many-job to single quantum platform} \added{(Fig~\ref{fig:qc_res:single})}, and
\emph{per-job to single quantum platform} \added{(Fig~\ref{fig:qc_res:per_job})}.
In the \emph{many-job to single quantum platform} model, multiple jobs can reserve the same quantum platform, which can be a classical machine simulating a quantum resource, or a physical quantum device.
In contrast, the  \emph{per-job to single quantum platform} model pairs each job to a quantum platform (real or simulated). 
This latter model requires a large set of quantum platforms, which is not currently realistic with physical devices, but we can exploit HPC based quantum simulation to implement the model.

\section{Framework Architecture}
\label{sec:framework}

\begin{figure}
    \centering
    \includegraphics[width=0.75\columnwidth]{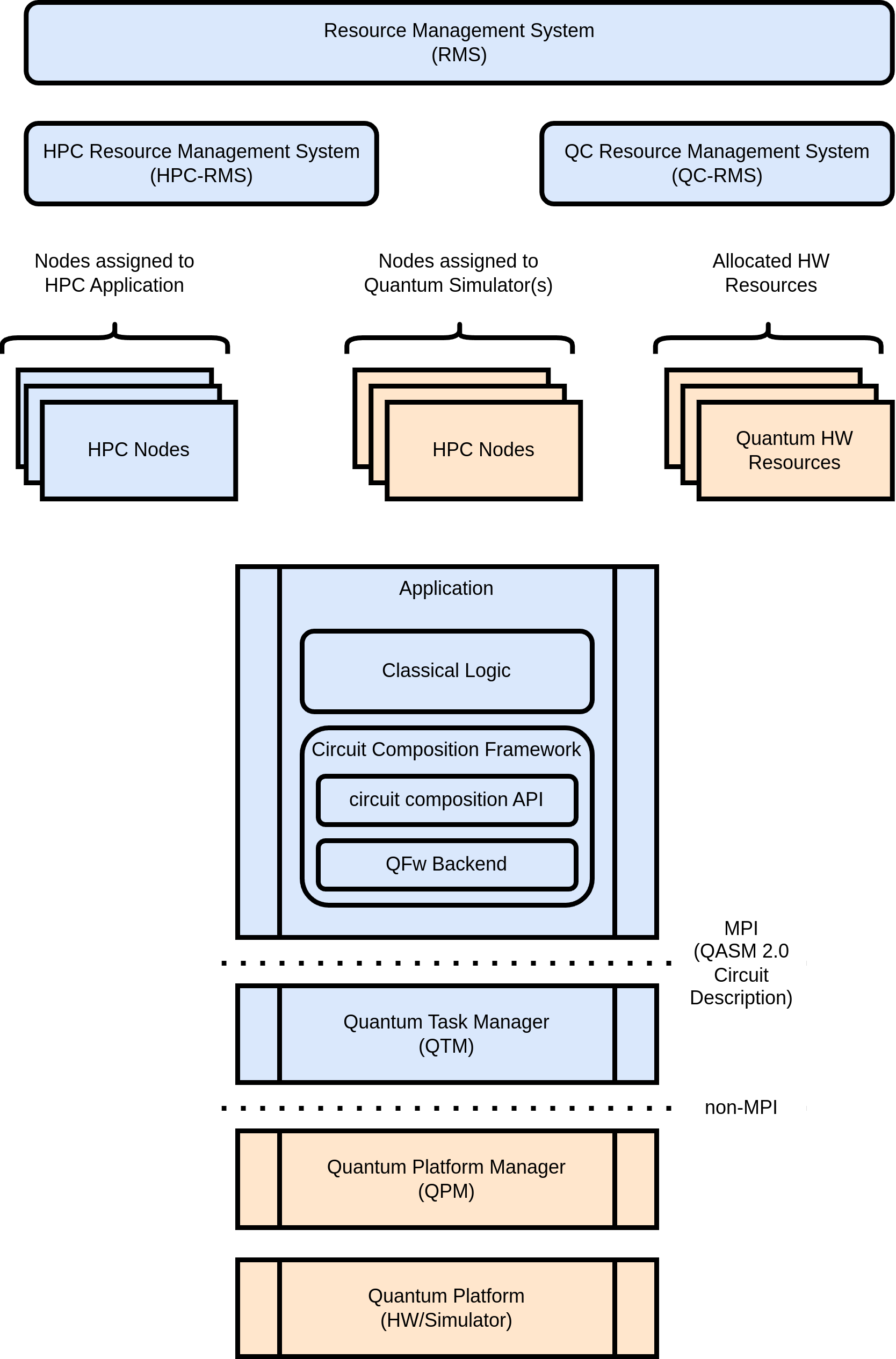}
    \caption{Architectural Overview}
    \label{fig:qfw_arch_overview}
\end{figure}

Given the HPC/QC integration spaces and models described in this paper, we developed a Quantum Framework (QFw) (Figure~\ref{fig:qfw_arch_overview}) aimed at providing researchers the ability to benefit from the full power of HPC resources. The QFw is also designed to allow hybrid applications to transition from executing with a simulation backend to executing on real hardware without requiring any changes.

In practice, applications which use the QFw, specify the HPC resources they require for the classical logic and quantum logic independently. They can then use any circuit composition software suited to their respective use cases. The QFw provides a backend for the conversion of native quantum circuit structures into QASM 2.0~\cite{openqasm2}. This common quantum task format is passed to lower layers of the framework for further processing.

The Quantum Task Manager (QTM) layer may apply specific workflows, such as circuit cutting for southbound tasks and aggregation for northbound results. The Quantum Platform Manager (QPM) manages communication with the underlying platform. It receives the quantum tasks from the QTM in standardized format and executes them through platform specific operations. The QPM provides a common API, which can have a platform dependent implementation. The framework provides common utility functions that  can be used by any QPM to streamline the implementation. The QFw can handle multiple QPMs to support the simultaneous utilization of different quantum platforms.

\section{Dynamic Simulation Environment}
\begin{figure}
    \centering
    \includegraphics[width=0.85\columnwidth]{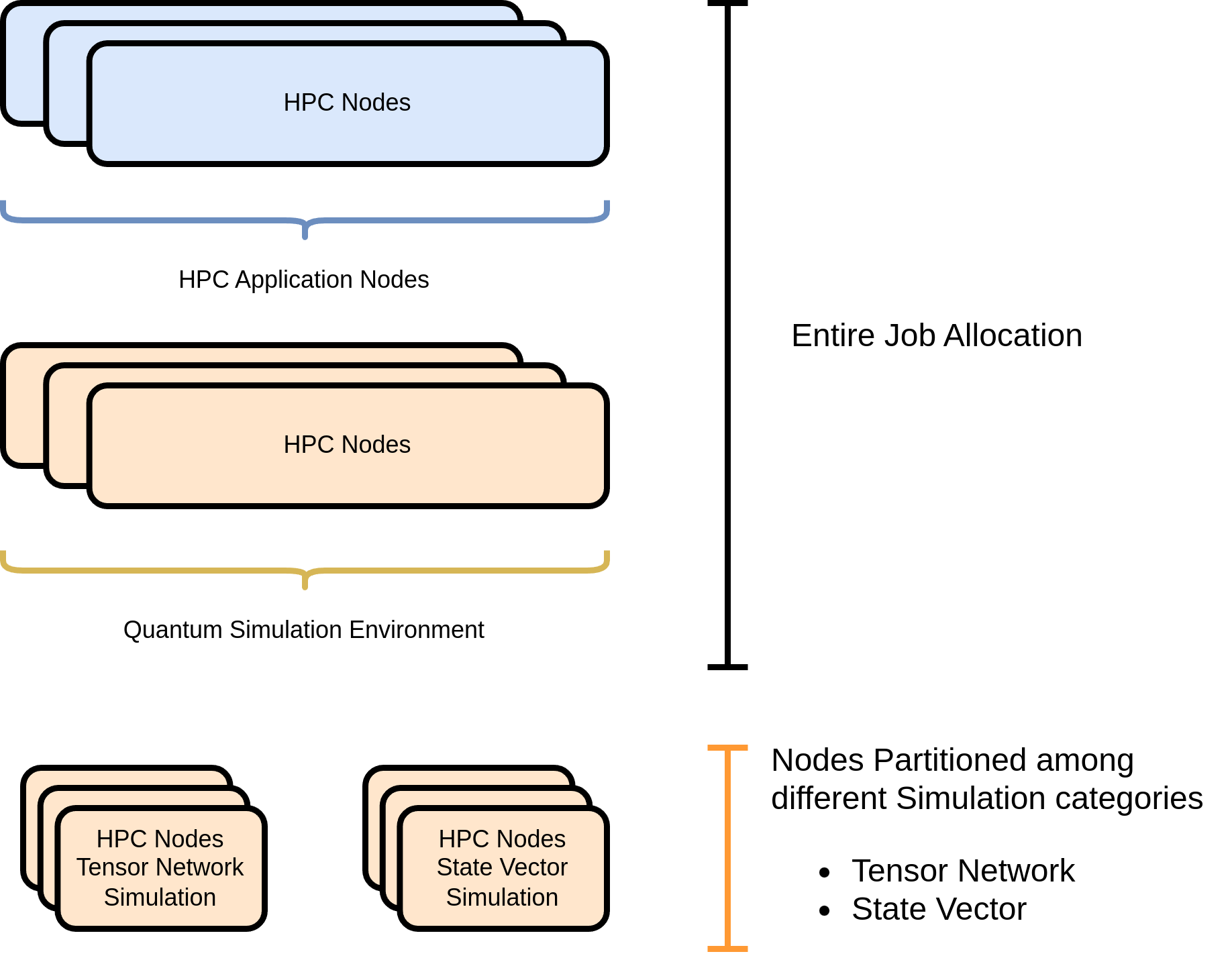}
    \caption{Dynamic Simulation Environment}
    \label{fig:dynamic_qsim}
\end{figure}
Even as fault-tolerant quantum computers materialize, quantum simulators will continue to be needed, due to the challenges in accessing quantum hardware, which is anticipated to be rare and costly. It is, therefore, a logical strategy to use simulators for rigorous testing and debugging of quantum programs prior to deployment on physical hardware.
To meet this demand, the QFw architecture includes the creation of a versatile
and expandable simulation environment, as illustrated in
Figure~\ref{fig:dynamic_qsim}, capable of accommodating diverse simulation
backends. This provides researchers the tools necessary to develop and verify their quantum algorithms in an error free environment.

The simulation environment is composed of a set of HPC nodes which are allocated concurrently with HPC nodes dedicated to the classical portion of the hybrid application. It can be partitioned based on user input into discrete segments responsible for executing various types of quantum simulators, such as tensor network simulators and state vector simulators. When an application submits quantum tasks for execution on the simulation environment, the environment employs heuristics—considering factors such as qubit count and gate depth—to determine the most appropriate simulators for the task. User preferences take precedence over default heuristics, allowing for customized simulation configurations.

Upon receipt of queued quantum tasks, the environment dynamically evaluates the necessary number of simulators and configures itself accordingly to efficiently handle concurrent simulations. These simulators can be spawned as MPI jobs, spanning multiple processes and nodes if necessary. This allows simulators to operate on large quantum tasks that otherwise could not be simulated. Conversely, for smaller quantum tasks that do not require multiple MPI processes, the environment can spawn multiple simulators on HPC resources to execute several quantum tasks concurrently.

This adaptive approach optimizes the utilization of HPC resources. Moreover, it gives researchers the tools necessary to experiment with different simulators. For instance, while some quantum tasks may benefit from tensor network simulators, others may require state vector simulators. Additionally, this approach allows applications to harness both types of simulators simultaneously, enabling tailored simulation strategies based on the specific needs of hybrid HPC/QC applications.

\section{Prototype}
\label{sec:prototype}

In order to verify the feasibility of the outlined design we developed a prototype to prove stated concepts. Although the prototype was tested on the \emph{Frontier} supercomputer, there is nothing in the design or prototype implementation preventing it from being deployed on other architectures and HPC environments.

The prototype encompasses the following key elements:
\begin{itemize}
    \item \textbf{Resource Allocation:} Allocating resources for both the HPC classical job and the simulation environment concurrently
    \item \textbf{Setting up the simulation environment:} Establishing the infrastructure for the simulation environment, enabling dynamic execution of multiple MPI jobs
    \item \textbf{Executing SupermarQ} \cite{supermarq}{\bf :} Generating circuits using SupermarQ, converting them to QASM 2.0, and queuing them on the simulation environment for execution
\end{itemize}

Since the Frontier software ecosystem uses SLURM as it's resource manager, we use SLURM's heterogeneous jobs~\cite{slurm:hetero} feature to allocate the HPC nodes required for both the HPC/QC hybrid application and the simulation environment. SLURM's heterogeneous jobs feature allows the allocation of both of these sets of resources concurrently. It presents these resources as two components of the same job. While heterogeneous jobs are not strictly necessary for a simulation environment comprised of classical HPC nodes, their utilization ensures a smoother transition to real quantum hardware. In scenarios involving actual hardware, SLURM's capability to allocate various resource types simultaneously within a unified job framework becomes imperative.

This can be done using sbatch as follows:

\begin{lstlisting}[frame=single, basicstyle=\small\ttfamily, breaklines=true]
# job component 1
#SBATCH -N <number of nodes>
#SBATCH -t 1:00:00

#SBATCH hetjob

# job component 2
#SBATCH -N <number of nodes>
\end{lstlisting}
or it can be dynamically allocated as follows

\begin{lstlisting}[frame=single, basicstyle=\small\ttfamily, breaklines=true]
salloc -N 1 -t <time limit> -A <account> : \
       -N 2 -t <time limit> -A <account>
\end{lstlisting}

Upon resource allocation, the system presents two job components: the first encompassing nodes dedicated to the hybrid HPC/QC application, and the second encompasses nodes allocated for the simulation environment. Subsequently, the prototype initiates a custom-built RPC framework named Distributed Execution Framework (DEFw), scripted in Python, on the initial node within the simulation environment.

DEFw initializes a resource manager, with which all other DEFw instances register. The Quantum Platform Manager (QPM), as outlined in the architectural overview, is implemented as a DEFw instance. Multiple QPMs can operate within the simulation environment, each managing an independent quantum platform and registering with the resource manager. In this context, a quantum platform pertains to a specific simulation category, such as tensor network simulator or state vector simulator. For simplicity, a single QPM instance runs in the simulation environment, assuming solely tensor network simulators are active.

The application utilizes DEFw mechanisms to access the QPM via the resource manager. DEFw provides an RPC implementation that enables the application to utilize an exported QPM API for queuing circuits for simulation. The QPM exports APIs which create circuits, executes circuits synchronously or asynchronously, and retrieves circuit results.

The prototype leverages
 eXtreme-scale ACCelerator~(XACC)~\cite{McCaskey_2020},
 Tensor Network Quantum Virtual Machine~(TNQVM)~\cite{Nguyen2021TensorNQ}, and
 the numerical tensor algebra library  ExaTensor~\cite{10.3389/fams.2022.838601}.
\added{TNQVM is a vital tool for the development, verification, and validation of hybrid quantum-classical algorithms on near-term quantum co-processors. It has been modernized to leverage exascale high-performance computing platforms, enabling the simulation of quantum circuits at large-scale. 
This updated version is integrated into the XACC.}

Developed at ORNL, this stack has been tailored to compile and link against the ROCm library for execution on the AMD-based \emph{Frontier} supercomputer. Furthermore, it has been compiled with MPI enabled to function as an MPI process, allowing TNQVM and ExaTensor to span multiple MPI processes and nodes. A straightforward C++ circuit runner program has been implemented. It links against the XACC, TNQVM, and ExaTensor libraries. This program accepts a QASM 2.0 circuit description as input and simulates it using TNQVM.

On Frontier, TNQVM was capped at 8~MPI processes per node.  This was based
on the Frontier node architecture~\cite{frontier:sc23} and an observation
that assigning multiple TNQVM MPI processes to a GPU exceeded GPU memory and
resulted in simulator failures.

A key aspect of the simulation environment is to dynamically manage the allocation and run multiple simulators at different scales, therefore fully utilizing the HPC allocation dedicated to the simulation environment. 
The prototype leverages the PMIx Runtime Environment~(PRTE), which is included with Open~MPI, to create a persistent Distributed Virtual Machine~(DVM) to manage processes.
Utilizing the Frontier-adapted Open~MPI~\cite{pritchard:cug23:ss11}, we execute the circuit runner as an MPI application. 
An additional argument ({\tt --dvm}) is added to the {\tt mpirun} command-line to reference the PRTE DVM server.
The single DVM server instance maintains a cohesive view of resource availability within the simulation environment.
All simulator instances are mapped to an L3 cache line and bound to the cores within ({\tt --map-by l3cache:pe=6 --bind-to core}). This gives each MPI simulator process multiple cores to use as well as ensures that each MPI process is bound to a unique GPU on the Frontier node. 

Subsequently, PRTE orchestrates multiple MPI runs, binding each MPI process and safeguarding against resource overutilization. 
For instance, in a scenario where the simulation environment is comprised of 2~nodes, 3~simulator instances can be run simultaneously. The first instance utilizes 4~MPI processes, all executed on the initial node.
The second instance employs 8~processes, with 4~running on the first node and the remaining 4 on the second node. 
Finally, the third instance utilizes 4~processes, all executed on the second node.

The mapping and the number of MPI processes to run per simulator instance is determined by the QPM. The QPM uses simple heuristics that considers the number of qubits in the circuit to determine the number of simulator MPI processes needed to simulate the circuit. More nuanced criteria for managing the mapping of quantum circuits to simulator MPI process remains a future endeavor. 

\added{In addition to TNQVM,  we created another QPM backend to support the
integration of NWQ-Sim \cite{li2021svsim,li2020density}. 
For each circuit execution, the application can optionally specify the simulator backend it wishes to use. The infrastructure can then dynamically launch the specified simulator, which allows for support of multiple simulators and their simultaneous execution.} 

\vspace{-0.2cm}

\section{Evaluation}
\label{sec:eval}

\added{
As part of our evaluation we configured QFw to use different 
frontends (Qiskit, PennyLane) 
and  backends (TNQVM, NWQ-Sim).  We tested using a synthetic benchmark tool and 
simple quantum applications (QAOA, Hamiltonian simulation).
} 

\added{SupermarQ \cite{supermarq} is a pioneering quantum benchmark suite designed to address the challenges of measuring and comparing performance across diverse quantum computing architectures.}
SupermarQ was used to
to generate a GHZ circuit, convert it to QASM~2.0 and then use the QPM API
to run the circuit. This application runs on the first job component
dedicated to the classical HPC application. However, it currently does not
run as an MPI process.

\begin{lstlisting}[frame=single, basicstyle=\small\ttfamily, breaklines=true]
ghz = supermarq.benchmarks.ghz.GHZ(num_qubits=x)
cir = ghz.circuit()
qasm = cir.to_qasm()
info = {}
info['qasm'] = qasm
info['num_qubits'] = x 
info['num_shots'] = 1 
info['compiler'] = 'staq'
cid = qpm_api.create_circuit(info)
rc, circ_result, stats = qpm_api.sync_run(cid)
\end{lstlisting}

For the first iteration of the prototype we chose time-to-solution as a way
to evaluate the usefulness of the prototype. SupermarQ was used to generate
a 20~qubit circuit. 
Briefly, the circuit initializes the $0th$ qubit via a Hadamard gate, the
remaining $n$ qubits are set with a CNOT of the $n-1$ qubit, and then
measurements read results to 20~classical registers.

The prototyped framework was used to simulate this circuit synchronously one
time ($52.47$ seconds) and then eight times concurrently ($66.97$ seconds).
The test used a basic mapping of one MPI process for each 10~qubits, such
that the 20~qubit circuit required two MPI processes for the simulator.
This highlights the time improvements for simulating independent circuits concurrently,
as well as the advantages of using a more intelligent resource management layer.
This also demonstrates the flexibility of using the presented framework to drive
experiment campaigns with different execution patterns and configurations with minimal
user effort.

\added{
Furthermore, we integrated a Qiskit application to solve a Max-Cut QAOA problem using QFw. 
We tested the application with both the TNQVM and NWQ-Sim backends, highlighting the flexibility of QFw in dynamically selecting the backend simulator. Additionally, by integrating another application that utilizes PennyLane, we demonstrated the capability to use any frontend with any backend simulator.
}

\section{Future Work}
\label{sec:future}

\added{The current work targets the on-premises model and is biased toward the use of HPC resources in support of quantum simulators. As new hardware resources become available, we intend to
incorporate those and refactor the QFw implementation as needed.  We intend
to further investigate work by other organizations (e.g.,
HPCQC~\cite{www:hpcqc}) that have
begun looking at similar problems of HPC/QC integration~\cite{schulz:accelerating:ieee22} to
see what aspects can be leveraged or unified.  Specifically, the
adaptability/malleability of workloads is something that these groups have
studied and it would be useful to incorporate these capabilities into our
framework implementation.
} 

\added{The QFw architecture supports the insertion of heuristics,
e.g., QPM layer provides ``hook'' points for inserting heuristics.  The
implementation provides basic proof-of-concept heuristics for the two
simulators used in the initial phase of development.  In future work it
would be useful to improve the set of heuristics, e.g., support more
application introspection methods using circuit size, resource load, etc.
It would also be interesting to explore new/alternate simulators and their
relevant heuristics.
} 

\added{We have further plans to modify an actual quantum hybrid applications to make use of the QFw prototype; thereby allowing it to run on Frontier at scale. The work required to bring the prototype to this stage will align it closer to the ultimate vision outlined by the architecture.
} 

\added{Finally, we plan to further enhance the framework.  The initial
implementation provides a platform for exploring the emerging space of quantum
enabled HPC applications.  For example, there are likely load imbalances
between the HPC (classical) and quantum compute phases of these applications
that can benefit from a more intelligent resource management layer.  We intend
to use QFw to investigate more application efficiency and resource utilization
trade-offs, e.g., do not let quantum resources sit idle if bulk of work is
classical with small quantum phases.  This is an area where the above mentioned
adaptability/malleability work would likely be useful.  The initial framework
provides a basis to begin gathering the necessary usage information to help
inform the resource allocation policies and explore usage patterns.
} 


\section{Conclusions}
\label{sec:conclusions}
The Quantum Framework~(QFw) provides users the tools they require to advance their quantum research on the Frontier supercomputer. Unlike other emerging tools in the field, it gives researchers the flexibility to use any frontend circuit building software stack suitable for their hybrid application with any backend simulation package. This reduced friction allows researchers to focus on their research tasks, without being hindered by the need to port the application to another framework. It also removes the technical know-how required for running their application code on Frontier.

Furthermore, the QFw provides the mechanisms necessary to expand simulations on HPC systems beyond what would normally be possible, while at the same time allowing hybrid applications to transitions to running on physical quantum hardware without requiring any changes. 

The QFw is versatile enough to allow the integration of different quantum platforms without requiring alterations to either the infrastructure or the application. While incorporating additional quantum platforms may require the development of new Quantum Platform Managers~(QPMs) to interface with them, this process entails minimal effort. Moreover, the QFw's ability to support multiple QPMs enables it to concurrently utilize different quantum platforms, which is of benefit to the application. This design reduces the effort needed to integrate simulated or physical quantum platforms. The ultimate objective is to use the QPM as a plugin architecture, which presents a common API any platform provider needs to implement in order to integrate with the QFw.

\replaced{The developed prototype was evaluated using SupermarQ, a Qiskit application and a PennyLane application. SupermarQ generates circuits for differing quantum algorithms using varying qubit counts and gate depths. QFw manages the simulation of these circuits on TNQVM or NWQ-Sim. To further illustrate the flexibility afforded by QFw, we integrated a Qiskit and PennyLane applications, which simulate their quantum circuit workload on the available backend simulators via the QFw. Consequently, we can construct a performance profile for the underlying simulator. The QFw architecture allows the integration of other simulators, allowing us to perform similar profiling and comparison.}
{The prototype we developed uses SupermarQ to generate circuits for differing quantum algorithms which use varying qubit counts and gate depths. It then simulates these quantum circuits using the TNQVM simulator. Consequently, we can construct a performance profile for the underlying simulator. Other simulators can be integrated under the QFw, allowing us to perform the same profiling on them as well.} 


\section{Acknowledgments}

\begin{footnotesize}
This research used resources of the Oak Ridge Leadership Computing Facility
at the Oak Ridge National Laboratory, which is supported by the Office of
Science of the U.S. Department of Energy under Contract No.
DE-AC05-00OR22725.
\textsl{Notice}: This manuscript has been authored by UT-Battelle, LLC under Contract No. DE-AC05-00OR22725 with the U.S. Department of Energy.  The publisher, by accepting the article for publication, acknowledges that the U.S. Government retains a non-exclusive, paid-up, irrevocable, world-wide license to publish or reproduce the published form of the manuscript, or allow others to do so, for U.S. Government purposes. The DOE will provide public access to these results in accordance with the DOE Public Access Plan (http://energy.gov/downloads/doe-public-access-plan). 
\end{footnotesize}

\bibliographystyle{IEEEtran}
\bibliography{main}

\begin{thebibliography}{10}
\providecommand{\url}[1]{#1}
\csname url@samestyle\endcsname
\providecommand{\newblock}{\relax}
\providecommand{\bibinfo}[2]{#2}
\providecommand{\BIBentrySTDinterwordspacing}{\spaceskip=0pt\relax}
\providecommand{\BIBentryALTinterwordstretchfactor}{4}
\providecommand{\BIBentryALTinterwordspacing}{\spaceskip=\fontdimen2\font plus
\BIBentryALTinterwordstretchfactor\fontdimen3\font minus
  \fontdimen4\font\relax}
\providecommand{\BIBforeignlanguage}[2]{{%
\expandafter\ifx\csname l@#1\endcsname\relax
\typeout{** WARNING: IEEEtran.bst: No hyphenation pattern has been}%
\typeout{** loaded for the language `#1'. Using the pattern for}%
\typeout{** the default language instead.}%
\else
\language=\csname l@#1\endcsname
\fi
#2}}
\providecommand{\BIBdecl}{\relax}
\BIBdecl

\bibitem{8123664}
K.~A. Britt, F.~A. Mohiyaddin, and T.~S. Humble, ``Quantum accelerators for
  high-performance computing systems,'' in \emph{2017 IEEE International
  Conference on Rebooting Computing (ICRC)}, 2017, pp. 1--7.

\bibitem{10.1117/12.2303824}
\BIBentryALTinterwordspacing
T.~S. Humble, R.~J. Sadlier, and K.~A. Britt, ``{Simulated execution of hybrid
  quantum computing systems},'' in \emph{Quantum Information Science, Sensing,
  and Computation X}, E.~Donkor and M.~Hayduk, Eds., vol. 10660, International
  Society for Optics and Photonics.\hskip 1em plus 0.5em minus 0.4em\relax
  SPIE, 2018, p. 1066002. [Online]. Available:
  \url{https://doi.org/10.1117/12.2303824}
\BIBentrySTDinterwordspacing

\bibitem{9537178}
T.~S. Humble, A.~McCaskey, D.~I. Lyakh, M.~Gowrishankar, A.~Frisch, and
  T.~Monz, ``Quantum computers for high-performance computing,'' \emph{IEEE
  Micro}, vol.~41, no.~5, pp. 15--23, 2021.

\bibitem{Preskill2018}
\BIBentryALTinterwordspacing
J.~Preskill, ``Quantum computing in the nisq era and beyond,'' \emph{Quantum},
  vol.~2, p.~79, Aug 2018. [Online]. Available:
  \url{https://doi.org/10.22331/q-2018-08-06-79}
\BIBentrySTDinterwordspacing

\bibitem{ALEXEEV2024666}
\BIBentryALTinterwordspacing
Y.~Alexeev and {\it et. al.}, ``Quantum-centric supercomputing for materials
  science: A perspective on challenges and future directions,'' \emph{Future
  Generation Computer Systems}, vol. 160, pp. 666--710, 2024. [Online].
  Available:
  \url{https://www.sciencedirect.com/science/article/pii/S0167739X24002012}
\BIBentrySTDinterwordspacing

\bibitem{Cicconetti_2023}
\BIBentryALTinterwordspacing
C.~Cicconetti, M.~Conti, and A.~Passarella, ``Service differentiation and fair
  sharing in distributed quantum computing,'' \emph{Pervasive and Mobile
  Computing}, vol.~90, p. 101758, Mar. 2023. [Online]. Available:
  \url{http://dx.doi.org/10.1016/j.pmcj.2023.101758}
\BIBentrySTDinterwordspacing

\bibitem{qchpc}
\BIBentryALTinterwordspacing
V.~Bartsch, G.~C. de~Verdière, J.-P. Nominé, D.~Ottaviani, D.~Dragoni, F.~M.
  Chayma~Bouazza, D.~Bowden, C.~Allouche, M.~Johansson, O.~Terzo,
  A.~Scarabosio, G.~Vitali, F.~Shagieva, and K.~Michielsen, ``{\rm
  $<$QC$|$HPC$>$ Quantum for HPC},'' 2021. [Online]. Available:
  \url{https://www.etp4hpc.eu/pujades/files/ETP4HPC_WP_Quantum4HPC_FINAL.pdf}
\BIBentrySTDinterwordspacing

\bibitem{elsharkawy2023integration}
A.~Elsharkawy, X.-T.~M. To, P.~Seitz, Y.~Chen, Y.~Stade, M.~Geiger, Q.~Huang,
  X.~Guo, M.~A. Ansari, C.~B. Mendl, D.~Kranzlmüller, and M.~Schulz,
  ``Integration of quantum accelerators with high performance computing -- a
  review of quantum programming tools,'' 2023.

\bibitem{frontier:sc23}
S.~Atchley, C.~Zimmer, J.~Lange, D.~Bernholdt, V.~Melesse~Vergara, T.~Beck,
  M.~Brim, R.~Budiardja, S.~Chandrasekaran, M.~Eisenbach, T.~Evans, M.~Ezell,
  N.~Frontiere, A.~Georgiadou, J.~Glenski, P.~Grete, S.~Hamilton, J.~Holmen,
  A.~Huebl, D.~Jacobson, W.~Joubert, K.~Mcmahon, E.~Merzari, S.~Moore,
  A.~Myers, S.~Nichols, S.~Oral, T.~Papatheodore, D.~Perez, D.~M. Rogers,
  E.~Schneider, J.-L. Vay, and P.~K. Yeung, ``{\rm Frontier: Exploring
  Exascale},'' in \emph{Proc. of the International Conference for High
  Performance Computing, Networking, Storage and Analysis}, ser. SC'23.\hskip
  1em plus 0.5em minus 0.4em\relax ACM, 2023, doi:10.1145/3581784.3607089.

\bibitem{openqasm2}
A.~W. Cross, L.~S. Bishop, J.~A. Smolin, and J.~M. Gambetta, ``Open quantum
  assembly language,'' 2017.

\bibitem{supermarq}
\BIBentryALTinterwordspacing
T.~Tomesh, P.~Gokhale, V.~Omole, G.~Ravi, K.~N. Smith, J.~Viszlai, X.~Wu,
  N.~Hardavellas, M.~R. Martonosi, and F.~T. Chong, ``{\rm SupermarQ}: A
  scalable quantum benchmark suite,'' in \emph{2022 IEEE International
  Symposium on High-Performance Computer Architecture (HPCA)}.\hskip 1em plus
  0.5em minus 0.4em\relax Los Alamitos, CA, USA: IEEE Computer Society, apr
  2022, pp. 587--603. [Online]. Available:
  \url{https://doi.ieeecomputersociety.org/10.1109/HPCA53966.2022.00050}
\BIBentrySTDinterwordspacing

\bibitem{slurm:hetero}
\BIBentryALTinterwordspacing
SchedMD, ``{\rm SLURM:} heterogeneous job support,'' 2024, last-visited:
  21-apr-2024. [Online]. Available:
  \url{https://slurm.schedmd.com/heterogeneous_jobs.html}
\BIBentrySTDinterwordspacing

\bibitem{McCaskey_2020}
\BIBentryALTinterwordspacing
A.~J. McCaskey, D.~I. Lyakh, E.~F. Dumitrescu, S.~S. Powers, and T.~S. Humble,
  ``{\rm XACC}: a system-level software infrastructure for heterogeneous
  quantum–classical computing*,'' \emph{Quantum Science and Technology},
  vol.~5, no.~2, p. 024002, feb 2020. [Online]. Available:
  \url{https://dx.doi.org/10.1088/2058-9565/ab6bf6}
\BIBentrySTDinterwordspacing

\bibitem{Nguyen2021TensorNQ}
\BIBentryALTinterwordspacing
T.~Nguyen, D.~I. Lyakh, E.~F. Dumitrescu, D.~Clark, J.~Larkin, and A.~J.
  McCaskey, ``Tensor network quantum virtual machine for simulating quantum
  circuits at exascale,'' \emph{ACM Transactions on Quantum Computing}, vol.~4,
  pp. 1 -- 21, 2021. [Online]. Available:
  \url{https://api.semanticscholar.org/CorpusID:233324255}
\BIBentrySTDinterwordspacing

\bibitem{10.3389/fams.2022.838601}
\BIBentryALTinterwordspacing
D.~I. Lyakh, T.~Nguyen, D.~Claudino, E.~Dumitrescu, and A.~J. McCaskey, ``{\rm
  ExaTN}: Scalable gpu-accelerated high-performance processing of general
  tensor networks at exascale,'' \emph{Frontiers in Applied Mathematics and
  Statistics}, vol.~8, 2022. [Online]. Available:
  \url{https://www.frontiersin.org/articles/10.3389/fams.2022.838601}
\BIBentrySTDinterwordspacing

\bibitem{pritchard:cug23:ss11}
H.~Pritchard, T.~Naughton, A.~Shehata, and D.~Bernholdt, ``{{\rm Open~MPI for
  HPE Cray EX Systems}},'' in \emph{Proceedings of the Cray User's Group~(CUG)
  Annual Technical Conference}, May 2023.

\bibitem{li2021svsim}
A.~Li and S.~Krishnamoorthy, ``{\rm SV-Sim}: Scalable pgas-based state vector
  simulation of quantum circuits,'' in \emph{Proceedings of the International
  Conference for High Performance Computing, Networking, Storage and Analysis},
  2021.

\bibitem{li2020density}
A.~Li, O.~Subasi, X.~Yang, and S.~Krishnamoorthy, ``Density matrix quantum
  circuit simulation via the bsp machine on modern gpu clusters,'' in
  \emph{Proceedings of the International Conference for High Performance
  Computing, Networking, Storage and Analysis}, 2020.

\bibitem{www:hpcqc}
``{HPCQC: Where Quantum accelerates the future of Supercomputing},'' 2024,
  \url{https://www.hpcqc.org}.

\bibitem{schulz:accelerating:ieee22}
M.~Schulz, M.~Ruefenacht, D.~Kranzlmüller, and L.~B. Schulz, ``{Accelerating
  HPC With Quantum Computing: It Is a Software Challenge Too},''
  \emph{Computing in Science and Engineering}, vol.~24, no.~4, pp. 60--64,
  2022.

\end{thebibliography}

\end{document}